\documentstyle[preprint,aps]{revtex}
\begin{document}
\draft
\title{Calculated optical properties of Si, Ge, and GaAs under hydrostatic
pressure} 
\author{M. Alouani and J. M. Wills }
\address{ Department of Physics, The Ohio State University, Ohio 43210-1368 }
\address{ Theoretical Division, Los Alamos National Laboratory
Los Alamos, New Mexico 87545}
\date{\today}
\maketitle

\begin{abstract}
The macroscopic dielectric function in the random-phase-approximation  
without local field effect has been implemented using the local density
approximation with an all electron, full-potential linear muffin-tin
orbital   basis-set. This method is used to investigate the optical 
properties of the semiconductors Si, Ge, and GaAs under hydrostatic
pressure. 
The pressure dependence of the effective dielectric function is compared
to the experimental data of Go\~ni and coworkers,  and an 
excellent agreement is found when the so called ``scissors-operator'' shift
(SOS) is used to account for the correct band gap at $\Gamma$. The effect
of the 
$3d$ semi-core states in the interband transitions hardly changes the static 
dielectric function, $\epsilon_\infty$; 
however, their contribution to the
intensity of absorption for higher  photon energies 
is substantial. 
The spin-orbit coupling 
has a significant effect on $\epsilon_\infty$ of Ge and GaAs, but not of
Si. The $E_1$ peak in the dynamical dielectric function is strongly
underestimated for Si, but only slightly for Ge and GaAs, suggesting that
excitonic effects might be important only for Si. 

\end{abstract}
\narrowtext
\section{INTRODUCTION}
\label{sec:intro}

The experimental determination of 
the optical properties of bulk semiconductors can now be obtained 
with high precision \cite{goni,cardona,aspnes},  yet our theoretical
understanding is far from  complete.
The static dielectric constant, which can be obtained from a functional
derivative of the electron density with respect to the total Kohn-Sham 
potential evaluated at the ground state, hence a ground state property, is
over estimated  by  the local-density approximation (LDA) calculation 
\cite{baroni,alouanib,zachary}. 
The inclusion of the gradient correction to the pseudopotential LDA 
reduces slightly the discrepancy in the case of Silicon \cite{dalcorso}.  
The underestimation of the $E_1$ peak and the overestimation of the 
$E_2$ peak of the imaginary part of the dielectric function, 
$\epsilon_2 (\omega)$, by one-electron band theory  have generated 
theoretical work for almost two decades to account for these discrepancies.
It was clear from the beginning  that including excitonic
effects, which have been detected experimentally \cite{pollak}, 
could remove some of the   disagreement   with experiment 
\cite{hanke,sham,hanke84,delca}.
However, the model calculations used to correct $\epsilon_2(\omega)$ 
have  produced only a qualitative understanding of the problem. In 
particular, the latest model by Hanke, Mattausch and Strinati 
based on the time-dependent screened Hartree-Fock approximation
and  including both the local field and the excitonic effects,
described correctly   the $E_1$ 
peak but underestimated significantly the $E_2$ peak of Si.  The reason
for the underestimation of $E_2$ was attributed to a bad representation 
of the band structure of Si  by their Slater-Koster 
parameterization \cite{hanke84}. Del
Castello-Mussot and Sham \cite {delca} based their latest calculation 
on a $\bf k.p$ 
model around the L point (where $E_1$ originates) and a multiple
plane-wave model around the X points (where  $E_2$ originates), and 
solved the Bethe-Salpeter equation containing the excitonic effect. Their
model is an improvement over the non-interacting approximation: 
the $E_1$ peak becomes stronger and the $E_2$ peak weaker. This model
is very promising, but,  being based on a  $\bf k.p$ 
approximation to the band structure, it provided only a 
qualitative correction to the intensities of the  $E_1$ and $E_2$
structures. Calculations ignoring excitonic effects, but  including 
the local-field effect, underestimated 
 both $E_1$ and $E_2$ peaks \cite{louie}.

One way to make theoretical 
progress in this field is to  determine the correct contribution of the
one-electron theory to the optical properties of semiconductors. This
allows us to define precisely the size of the many 
body corrections to the one electron theory. However, a common belief
these days is that the eigenvalues and vectors of the Kohn-Sham (KS) 
equations \cite{kohn}
have no direct physical meaning and hence should not be used to calculate
optical spectra of materials. Only ground state properties derived from
the total energy as a function of the electron density have in principle
a direct physical meaning.

While LDA was indeed intended to  calculate ground state properties it 
could also be viewed as a simplified quasi-particle (QP) theory 
where the self-energy is local and static 
($\Sigma({\bf r}, {\bf r^\prime}, t) \approx V_{xc} ({\bf r})
\delta ({\bf r} -  {\bf r^\prime} ) \delta (t)$, here $V_{xc} ({\bf r})$
is the local exchange and correlation potential as, for example,  
parameterized by Von Barth and 
Hedin \cite{vbarth}.  The KS  eigenvalues are 
then QP energies and could be compared to experimental data. 
This argument
is supported by calculations using the GW approximation of Hedin\cite{hedin}.
These calculations   showed that the valence QP energies
 of semiconductors are in good agreement with LDA and the conduction QP 
energies differ by approximately a rigid energy shift \cite{hybertsen,gss}. 
In the literature this
shift is often called ``scissors-operator'' shift (SOS) \cite{zachary}. 

First-principles local density approximation calculations started more than 
two decades ago, but the major problem of LDA, beside  the well understood 
 energy band gap problem\cite{plevy}, is the numerical difficulty to 
determine selfconsistent electronic structure and optical matrix
elements  using a complete basis-set. 
The early ab-initio calculation of the optical properties of
semiconductors by Wang and Klein, using a selfconsistent linear combination
of gaussian orbitals  produced  static dielectric functions in good 
agreement with experiment\cite{wangklein}. But this agreement is 
fortuitous because the
band gaps produced by this method are much larger than the LDA band gaps.
The recent calculations of 18 semiconductors by Huang and Ching using 
an orthogonalized linear combination of atomic orbitals method produced 
LDA static dielectric functions which are in general smaller than experiment
despite the fact that their band gaps are much larger than the all-electron
or pseudopotential LDA band gaps \cite{huang}. 
Those underestimated static dielectric constants are most likely due to the 
incompleteness of the basis-set used in their calculations. 

Most of the theoretical studies of the optical properties of semiconductors 
in the literature
use several approximations within the LDA, ranging 
from the use of spherical potentials \cite{alouanib} to the use of 
pseudopotentials \cite{zachary} instead of all electron LDA potentials. 
In this paper we report precise calculations of  the optical properties of bulk 
semiconductors Si, Ge, and GaAs under hydrostatic pressure using an 
all electron LDA linear muffin-tin orbital basis-set\cite{oka}, in which  
no shape
approximation is made for either the potential or the charge density
\cite{wills}. 
The semi-core $3d$ of Ga and Ge are included in a fully hybridizing  valence
basis set, and the rest of the core states are
allowed to relax selfconsistently. The effect of spin-orbit coupling is also 
investigated.  A systematic 
check of the $f$-sum rule is performed for all the calculations. We hope that 
this accurate LDA calculation will provide an excellent starting point for 
the determination of the local field and the  excitonic 
effects in the optical spectra of semiconductors 
\cite{hanke,sham,hanke84,delca,louie}.

We  have found that the static dielectric function,
$\epsilon_\infty$, which is a ground state property, is overestimated   
by LDA over all pressure range, and that an    
excellent agreement with the experimental results of Go\~ni and coworkers \cite{goni} 
for $\epsilon_\infty$ of GaAs and Ge under hydrostatic pressure is achieved only
when  the  so-called scissors-operator shift is used to  account for the 
correct band gap at $\Gamma$.  
The inclusion of the  $3d$ semi-core states
of Ge and GaAs in the interband transition has almost no significant
effects in $\epsilon_\infty$; however, the $3d$ interband transitions  
contribute significantly to the magnitude  of $\epsilon_2 (\omega) $ above
25 eV for Ge, and above 12 eV for  GaAs. The spin-orbit 
coupling  increases the LDA values by about few
percents. 

The rest of the paper is organized as follows. In section II, we describe the
method of calculation of electronic structure and the macroscopic 
dielectric function based on our
all-electron full-potential LMTO basis-set. In section III we present the
electronic properties of Si, Ge, and GaAs and compare them to existing 
theoretical results. The calculated dielectric functions and a discussion
 about  including the semi-core states and the spin-orbit
coupling will be presented in section IV. In the same section we also 
compare our
static dielectric function under hydrostatic pressure with the experimental 
results of Go\~ni {\it et
al.} \cite{goni}. The conclusion is given in Sec. V.

\section{METHOD OF CALCULATION}

\subsection { \bf All electron full-potential wave function }
\label{sec:Wave function of the crystal in full-potential LMTO:}

The full-potential linear muffin-tin orbital method
in its scalar-relativistic
and full-relativistic forms \cite{wills} is used here to
calculate the electronic structure and the optical properties of Si, Ge,
and GaAs under hydrostatic pressure. The Kohn-Sham \cite{kohn} equations
are solved for a general potential without any shape approximation
\cite{wills}. In this subsection we describe the Bloch wave function 
inside the so-called muffin-tin spheres and the interstititial region.
A correct determination of the crystal wave function is necessary for
the accurate determination  of the optical matrix elements.

As for  the cellular methods, the space is divided into non-overlapping
muffin-tin spheres surrounding atomic sites where the Schr\"odinger or the
 Dirac equation 
for each principle quantum number $\nu$ and momentum channel $\ell$ is
solved for a fixed  energy $E_{\nu\ell}$. In these muffin-tin spheres 
the trial wave function 
is linearized in terms of the solution of Schr\"odinger equation 
$\phi_{\tau\ell}$ and its energy derivative $ \dot \phi_{\tau \ell}$  
for the  energy $E_{\nu\ell}$,  and  for an atom of type $\tau$ and 
momentum channel $\ell$ \cite{oka,commenta}.


It can be shown  that the Bloch wave function of site $\tau$ calculated at site
$\tau^\prime$ in the unit cell of the crystal  at 
${\bf R} = {\bf 0} $ is given by \cite{wills}:
\begin{equation}
\chi_{\tau \ell m}^{\bf k} ({\bf r})|_{\tau^\prime} =
\sum_{\ell^\prime m^\prime}
\phi_{\tau^\prime\ell^\prime m^\prime}({\bf r - \tau^\prime})
B^{(1)\tau\tau^\prime}_{\ell^\prime m^\prime,\ell m}(\kappa,{\bf k})
+ \dot\phi_{\tau^\prime \ell^\prime m^\prime}({\bf r - \tau^\prime})
B^{(2)\tau\tau^\prime}_{\ell^\prime m^\prime,\ell m}(\kappa,{\bf k})
\label{wavemuffin}
\end{equation}
Where $B^{(1)\tau\tau^\prime}_{\ell^\prime m^\prime,\ell m}(\kappa,{\bf k})$
and $B^{(2)\tau\tau^\prime}_{\ell^\prime m^\prime,\ell m}(\kappa,{\bf k})$ are
renormalized structure constants obtained from the crystal structure
constants $B^{\tau\tau^\prime}_{\ell^\prime m^\prime,\ell m}(\kappa,{\bf
k})$  to ensure that the Bloch wave function is
continuous and differentiable at the boundary of each muffin-tin sphere. 

In the interstitial region, the muffin-tin orbitals are 
spherical wave solutions  $H_\ell$ to the Helmholtz equation with non-zero 
kinetic energy; these bases are Hankel
functions for negative kinetic energies or Neumann functions for positive
kinetic energies $ \kappa^2 $.
Such that each partial wave inside the muffin-tin
sphere is allowed to have different kinetic energy, $\kappa^2$, in the 
interstitial region. In this region the Bloch wave function is given by:
\begin{equation}
 \chi_{\tau \ell m}^{\bf k} ({\bf r}) = \sum_{\bf R} e^{i{\bf kR}}
 H_{\ell}(\kappa,|{\bf r-\tau - R}|) i^{\ell} Y_{\ell m} 
({\widehat{\bf r-\tau -R}})
\label{eqhankel}
\end{equation}
The interstitial-region Bloch function is expressed in plane waves
over the reciprocal lattice using Fourier transform : 

\begin{equation}
\chi_{\tau \ell m}^{\bf k} ({\bf r}) = \sum_{\bf G} f_{\bf K} ({\bf
k+G}) e^{i({\bf k+G}){\bf r}}
\label{planewave}
\end{equation}
Where ${\bf K}= \{\tau, \ell, m, E_{\ell}, \kappa\}$, 
here the parameter $E_{\ell}$ is the
linearization energy of the wave function in the muffin-tin sphere for the
$\ell$ momentum channel, $m$ is the  azumutal quantum number, and
$\kappa$ is the variational parameter whose  square is the 
kinetic energy in the interstitial region. The Fourier coefficients, 
$f_{\bf K}$, are obtained from a pseudo-wave function that is equal to the
crystal wave function in the interstitial region and represented by a
smooth function inside the muffin-tin spheres. The exact shape of these
pseudo-functions inside the muffin-tin spheres is not important. The only 
requirement is that they are continuous and differentiable at the sphere 
boundary and have zero slop at the origin of each sphere. 
The plane wave expansion is multiplied by a three dimensional step
function so that the wave function is kept only in the interstitial region. 
The knowledge of the Bloch wave function in the whole unit cell allows us
to calculate the Hamiltonian and overlap matrix elements in order to solve
the effective one electron Schr\"odinger equation.  

Three different kinetic energies were used for each subset of
{\it s} and {\it p} derived bases in the basis set; two kinetic energies were
used for bases derived from orbital parameters $\ell \geq 1$.  The basis sets
used in calculating total energies and structural properties were for Si:
 3(3{\it s}3{\it p}), 2(3{\it d}), for Ge:  2(3{\it d}), 3(4{\it s}4{\it p}),
2(4{\it d}), and for GaAs:  2(Ga 3{\it d}), 3(Ga 4{\it s}4{\it p}), 2(Ga 4{\it
d}), 3(As 4{\it s}4{\it p}), 2(As 4{\it d}); the pre-multiplicities in this
notation refer to the number kinetic energies used in this basis subset.  The
basis functions for each material comprised a single, fully hybridizing basis
set.  Note the presence of both 3{\it d} and 4{\it d} derived bases on Ga and
Ge.  A useful feature of the method used in these calculations is the ability to
incorporate basis functions derived from the same orbital atomic quantum numbers
but different principal atomic quantum numbers in a single fully hybridizing
basis set.  This feature entails the use of multiple sets of radial functions to
represent bases with different principle atomic quantum numbers.  This
capability was particularly useful in calculating the high-lying energy bands
which were used to obtain the dielectric functions to high energy; the basis
sets employed for this purpose are given in Table I.  Seven to eight kinetic 
energies were used in
the basis sets.  Accurate resolution of the bands to high energy was necessary
to converge the calculation of the real part of the dielectric function, which
was obtained from the imaginary part through the Kramers-Kronig relation.
An interesting consequence
of the relaxation of the Ga 3{\it d} states as valence states is a significant
decrease in the calculated band gap \cite{christensen}. 

For the core charge density, the Dirac equation is solved
selfconsistently, e.g., no frozen core approximation is used. 
The exchange and correlation potential is treated within the Von
Barth and Hedin parameterization \cite{vbarth}.
To account for the  relativistic effects in the dielectric function, the 
full-selfconsistent relativistic band structure is produced by 
including  the spin-orbit coupling to the Hamiltonian. 
In Table I we show
the orbitals used to describe the electronic states of Si, Ge, and GaAs.
This large number of orbitals is necessary to calculate accurately the 
eigenvalues and eigenvectors up to 5 Ry above the highest valence
states. These electronic states will be needed to determine the dynamical
dielectric function and the converged static dielectric function through
the use of Kramers-Kronig relations.

The completeness of basis-set, with different variational $\kappa$ values for
each partial wave in the interstitial region together with the Fourier 
representation allows the method to treat open 
structures like the zinc-blende structure studied here  without having to
resort to the so-called empty spheres \cite{christensen84}. The high energy 
 states are also determined more accurately due to the use of many
$\kappa$ values. 
As a test we show in Table II the eigenvalues of Si at high symmetry points
of the Brillouin zone compared with some recent results from first principle
calculations based on ab-initio pseudopotential and Gaussian orbital methods
\cite{cvl,germans}. The agreement of our calculation with the previous
calculations is  excellent. 
\label{sec:method}

\subsection { \bf Dielectric Function }

Here  we give a concise review of the determination
of the dielectric function of a semiconductor crystal due to the
application of an electric field. We also determine the approximations
used to obtain numerical results for Si, Ge, and GaAs under hydrostatic
pressure with or without scissors-operator shift.

A perturbative electromagnetic field of frequency $\omega$, and  a wave
vector $\bf q+G$ on a crystal produces a response of frequency $\omega$ and 
a wave vector  $\bf q+G^\prime$  (${ \bf G} $ and ${\bf G^\prime}$ being 
reciprocal lattice vectors). The microscopic field of wave vector 
 $ {\bf q+G^\prime}$ is produced by the umklapp processes as a result of the 
applied field  $E_0({\bf q + G}, \omega)$

\begin{equation}
 E_0({\bf q + G}, \omega) =
   \sum_{\bf G^\prime} \epsilon_{\bf G, G^\prime} ({\bf q}, \omega) E({\bf
q + G^\prime}, \omega) 
\label{field}
\end{equation}
\noindent
where  $ E({\bf q + G}, \omega)$ is the total field which produces the 
non-diagonal elements in the microscopic dielectric function   
 $\epsilon_{\bf G, G^\prime} ({\bf q}, \omega)$.
In the random phase approximation the microscopic dielectric function is 
given by \cite{adlerwiser}:

\begin{eqnarray}
\epsilon_{\bf G, G^\prime} ({\bf q}, \omega) & = 
\delta_{\bf G, G^\prime} - { 8 \pi e^2 \over 
{\Omega {|{\bf q + G}||{\bf q + G^\prime}|}}} 
\sum_{{\bf k}, n,n^\prime} {{f_{n^\prime,{\bf k+q}} -
f_{n,{\bf k}}} \over {E_{n^\prime,{\bf k+q}} - E_{n,{\bf k}}-\hbar \omega 
+ i\delta}} \langle n^\prime, {\bf k+q}| e^{i ({\bf q + G}) {\bf r}} | n,{\bf k} \rangle
\nonumber \\
& \times \langle n,{\bf k}| e^{-i ({\bf q + G^\prime}) {\bf r}}
 | n^\prime,{\bf k + q} \rangle  
\label{adler}
\end{eqnarray}

Here $n$ and  $n^\prime$ are the band indexes, $f_{n,{\bf k}}$ is the
zero temperature  
Fermi distribution, and $\Omega$ is the cell volume.  
The energies  $E_{n,{\bf k}}$ and the the crystal
wave function  $ | n,{\bf k} \rangle$ are produced for each band index $n$
and  for each wave vector  ${\bf k}$ in the Brillouin zone. 

The macroscopic dielectric function in the   infinite wave 
length limit is given by the inversion of the microscopic dielectric 
function:

\begin{eqnarray}
 \epsilon (\omega) = \lim_{{\bf q} \to {\bf 0}} { 1 \over
[\epsilon_{\bf G, G^\prime}^{-1} ({\bf q}, \omega)]_{\bf 0,0} } \\ \nonumber
 &    = \epsilon_{0,0}(\omega) -  \lim_{{\bf q} \to {\bf 0}} 
\sum_{{\bf G, G^\prime \ne 0}} \epsilon_{0,\bf G}({\bf q}, \omega)
T^{-1}_{\bf G, G^\prime} ({\bf q}, \omega) \epsilon_{\bf G^\prime,0}({\bf q},
\omega) \;
\label{localfield}
\end{eqnarray}

Where $T^{-1}_{\bf G, G^\prime}$ is the inverse matrix of $T_{\bf G,
G^\prime}$ containing the elements $\epsilon_{\bf G, G^\prime}$ with
${\bf G}$ and $\bf  G^\prime \ne \bf 0$. The first term of this 
equation is the interband contribution to the macroscopic dielectric 
function  and the second term represent the local-field correction to 
$\epsilon$. The most recent  ab-initio pseudopotentials calculation found 
that the local-field effect reduces the static dielectric function by at most 
5\% \cite{zachary}. Previous calculations with the same method have also
found a decrease of $\epsilon_{\infty}$ by about the same percentage
\cite{baroni,hybertsen}. We are looking at the effect of the local field
using our all-electron basis-set; it should be of interest to  compare
all electron results with these obtained using  the  pseudopotential method.
 
For insulators the dipole approximation of  the imaginary part of the 
first term 
of equation (\ref{localfield}) is given by \cite{eh59}: 

\begin{equation}
\epsilon_2(\omega)={   e^2 \over 3\omega^2\pi}\sum_{n,n^\prime}
\int  d{\bf k} | \langle n, {\bf k}| {\bf v}  | n^\prime, {\bf k}\rangle |^2 
f_{n,{\bf k}} (1-f_{{n^\prime},{\bf k}} ) \delta (e_{{\bf k},n^\prime,n} 
 -\hbar\omega)\;,
\label{ehrenreich}
\end{equation}

\noindent
Here  ${\bf v}$ is the velocity operator, and in the LDA ${\bf v} = {\bf
p }/m$ ($\bf p $ being the momentum operator), and 
where  $e_{{\bf k},n,n^\prime} = E_{n^\prime,{\bf k}} -
E_{{n},{\bf k}}$.
The matrix elements  $\langle n {\bf k}| {\bf p} |n^\prime {\bf  k}\rangle$  
are calculated for each projection $p_j={\hbar \over i}\partial_j$,
$j= x$ or $y$ and $z$, with the wave function $| n {\bf k}>$ expressed in terms
 of the full-potential LMTO  crystal wave function described by 
equations (\ref{wavemuffin}) and (\ref{planewave}).  The {\bf k}-space integration is performed using
the tetrahedron method \cite{ja72} with 480 irreducible {\bf k} points the 
whole Brillouin zone. 
The irreducible {\bf k}-points are obtained from a shifted  {\bf k}-space grid 
from the high
symmetry planes and $\Gamma$ point by a half step in each of the $k_x$,
$k_y$, and $k_z$ directions. This scheme  produces highly accurate integration 
in the Brillouin zone by avoiding high symmetry points.

To calculate these matrix elements we first defined a tensor operator of order 
one out of the momentum operator
 $\nabla_0 = \nabla_z = { \partial \over \partial z} \quad\hbox{and}\quad
 \nabla_{\pm 1} = \mp {1 \over \sqrt 2} ({ \partial \over \partial x}
\pm i {\partial \over \partial y} ) $.
The muffin-tin part of the momentum matrix elements is calculated using
the commutator $[ \nabla^2, x_\mu]=2 \nabla_{\mu} $ so that:
\begin{eqnarray}
\int_{S_{\tau}} d{\bf r} \phi_{\tau\ell^\prime} (r) & Y_{\ell^\prime m^\prime}
(\widehat{{\bf r -\tau}}) 
\nabla_{\mu} \phi_{\tau\ell}(r) Y_{\ell m}(\widehat{{\bf r-\tau}}) =
-{i \over 2} G^{1\mu}_{\ell m, \ell^\prime, m^\prime}\nonumber  \\
& \int_0^{S_{\tau}} r^2 dr \phi_{\tau\ell^\prime} ({2\over r} {d \over {d r}} r 
+ { {{\ell (\ell +1)} - {\ell^\prime (\ell^\prime +1)} } \over r }) 
\phi_{\tau\ell} (r)
\label{pmuffin}
\end{eqnarray}
where $G^{1\mu}_{\ell m, \ell^\prime, m^\prime}$ are the usual Gaunt
coefficients, and $S_{\tau}$ is the radius of the muffin-tin sphere of 
atom $\tau$.
 In the interstitial region the plane-wave
representation of the wave function (see equation ~\ref{planewave}) 
makes the calculation straightforward, but a  
special care has to be taken for the removal of the extra 
contribution in the muffin-tin spheres. However, 
 we find it much easier and faster to transform the
interstitial matrix elements as an integral over the surface of the 
muffin-tin spheres
using the commutation relation of the momentum operator and the
Hamiltonian in the interstitial region. The calculation of the 
interstitial momentum matrix elements  is then similar to the 
calculation of the interstitial  overlap matrix elements\cite{wills}. The 
$\kappa = 0$ case has been already derived by Chen using the Korringa, 
Kohn and 
Rostoker Green's-function method \cite{chen}. We have tested that both the
plane-wave summation and the surface integration provide the same results.

Equation (\ref{ehrenreich}) can not be used directly to determine the 
optical properties
of semiconductors, when the GW approximation or the scissors operator is 
used to determine the electronic structure. The velocity 
operator should be obtained from the effective momentum operator ${\bf
p^{eff}}$ which is calculated using the self-energy operator, 
$\Sigma({\bf r, \bf p})$, of the system \cite{delsole}:
\begin{equation}
{\bf v} = {\bf p^{eff}} /  m  = {\bf p} / m + \partial
\Sigma({\bf r, \bf p}) / \partial {\bf p}
\label{velocity}
\end{equation}
 GW calculations show  that  the quasiparticle wave function is 
almost equals  to the LDA wave function \cite{hybertsen,gss}.
Based on this assumption,
Del Sole and Girlanda show that the effective momentum operator 
${\bf p^{eff}}$ can be written in terms of the momentum operator ${\bf
p}$ as follows \cite{delsole}:
\begin{equation}
\langle n^\prime, {\bf k}|{\bf p^{eff}} | n, {\bf k}\rangle  =
 \langle n^\prime, {\bf k}| {\bf p}  | n, {\bf k}\rangle  
{e^{QP}_{{\bf k},n^\prime,n} / e_{{\bf k},n^\prime,n}},
\label{pqp}
\end{equation}
where $e^{QP}_{{\bf k},n^\prime,n} = 
E^{QP}_{n^\prime,{\bf k}} -  E^{QP}_{{n},{\bf k}}$
is the difference between the quasiparticle energy $E^{QP}_{n^\prime,{\bf
k}}$ of the unoccupied
state $|n^\prime, \bf k \rangle$ and the occupied state $|n, \bf k\rangle$. 
By substituting
Equation ~\ref{pqp} into equation ~\ref{ehrenreich}, it can be 
easily shown \cite{delsole} that in the case 
of the scissors operator, where all the empty states are shifted 
rigidly by  a constant energy $\Delta$, the imaginary part of the dielectric 
function is a simple energy shift of the LDA dielectric 
function towards the high energies by an amount $\Delta$, i.e., 
$\epsilon_2^{QP}(\omega) = \epsilon^{\rm LDA}_2(\omega -\Delta/\hbar)$. The 
real part of the dielectric function is then obtained from the 
shifted $\epsilon_2$  using Kramers-Kronig relations. 
The expression of $\epsilon_{\infty}^{QP}$ is given by:  
\begin{equation}
\epsilon^{QP}_{\infty}= 1 + {2e^2 \over   3\omega^2\pi^2}\sum_{n,n^\prime}
\int  d{\bf k}f_{n,{\bf k}} (1-f_{{n^\prime},{\bf k}} )
{ | \langle n, {\bf k}| {\bf p}  | n^\prime, {\bf k}\rangle |^2 \over 
 (e_{{\bf k},n^\prime,n} +\Delta) e_{{\bf k},n^\prime,n}^2}\;,
\label{estaticqp}
\end{equation}
$\epsilon^{QP}_{\infty}$ is very similar to $\epsilon^{LDA}_{\infty}$
except that one of the interband gap $e_{{\bf k},n^\prime,n}$ is
substituted by the QP interband gap $e_{{\bf k},n^\prime,n} +\Delta$. 

To test for the
accuracy of the calculation within the LDA the f-sum rule:
\begin{equation}
{2 \over 3 m n_v} \sum_{\bf k} \sum_{n,n^\prime} f_{n,{\bf k}}
(1-f_{{n^\prime},{\bf k}} ){|\langle n, 
{\bf k}|{\bf p} | n^\prime, {\bf k}\rangle|^2 \over e_{{\bf k},n^\prime,n}} 
= 1, 
\label{fsumrule}
\end{equation}
where $n_v$ is the number of valence bands,  is checked in all the 
calculations, and it     is
satisfied to within a few percent.

It is easily seen that the dielectric function $\epsilon_2^{QP}$ 
calculated using
the scissors-operator shift does not satisfy the 
sum rule ($\omega_P$ is the free-electron plasmon frequency):
\begin{equation}
\int_0^{\infty} \omega
\epsilon_2 (\omega ) d\omega  = {\pi \over 2} \omega_P^2
\label{intsumrule}
\end{equation}
because (i)
$\epsilon_2^{LDA}$ satisfies this rule, and (ii) $\epsilon_2^{QP}$ is
obtained by a simple shift of $\epsilon_2^{LDA}$ by the scissors-operator
$\Delta$ towards higher energies. 
Using the expression of the
quasiparticle dielectric function in the scissors-operator shift approximation we show 
that $\epsilon_2^{QP}$ satisfy the following integral sum rule:
\begin{equation}
\int_0^{\infty} \omega
\epsilon_2^{QP} (\omega ) d\omega  = {\pi\over 2} {\omega^{\prime}_P}^2 
\label{intsumruleqp}
\end{equation}
where   ${\omega^{\prime}_P}^2 = \omega_P^2 + 
{2 e^2 \Delta \over {3\pi^2 m^2}} \sum_{n, n^\prime} 
\int d{\bf k}{ | \langle n, {\bf k}| {\bf p}  | n^\prime, {\bf k}\rangle |^2 / 
e_{{\bf k},n^\prime,n}^2} f_{n,{\bf k}} (1-f_{{n^\prime},{\bf k}} )$. 
We recover the usual sum rule when  $\Delta $ is equal to zero. The non
simultaneous satisfaction of both the f-sum rule and the integral sum
rule given by Eq. 13 within the scissors approximation shows the  
limitation of this approximation. While the scissors operator
approximation describes nicely the low lying excited states, which is seen
in the good determination of the static dielectric function and the low
energy structures, i.e. $E_1$ and $E_2$, in the imaginary part of the dielectric
function, it seems to fail for the description of the higher excited states. 
This is not surprising because  the higher excited states which are free
electrons like are most probably well described by LDA and  need no scissors
operator shift.  This is supported by the fact that the the energy-loss 
function, -Im$\epsilon^{-1}$,  within the
LDA has it maximum roughly at  the free electron 
plasmon frequency whereas within the scissors approximation its maximum is
shifted to higher energies as given by equation (\ref{intsumruleqp}). 
Fig.~\ref{energyloss}
show the energy-loss function of GaAs calculated within LDA (full curve)
and within the scissors approximation (dashed curve). It is clearly seen
that the maximum of the LDA curve has a maximum which is closer to the 
free valence electron plasma frequency of 15.5 eV.
It is of general interest to see
whether the calculated dielectric function within the GW approximation 
satisfies the integral sum rule. For our purpose the scissors-operator
shift remains a good approximation for the description of the low-lying 
excited states of semiconductors and their optical properties.

\section{ELECTRONIC STRUCTURE OF Si, Ge, and GaAs}

The electronic structure of Si, Ge, and GaAs are obtained by solving the 
LDA equations by means of a full-potential LMTO 
basis-set as described above.
Table I shows  the orbitals used to describe the valence and
conduction bands during the  selfconsistency. The large number of orbitals
used is found necessary to obtain converged excited states up to 5 Ry
above the top of the valence states. However the total energy is 
insensitive to these high energy orbitals, but the presence of the
3$d$-core states of Ge and GaAs are  important \cite{christensen}.

Table II compares our band structure  of Si for some high symmetry  
points with some recent results from first-principles 
calculations based on pseudopotential and Gaussian orbitals methods
\cite{cvl,germans}. We found a  good agreement between our results and 
these calculations. This reflects the high accuracy of our
unoccupied states which are used to determine the dynamical dielectric 
function. 

Table III shows the calculated equilibrium structural parameters, i.e., 
the electronic pressure and the bulk modulus at the experimental  
unit cell volume, $V_0$,  and calculated 
cell volume, $V$. The calculated equilibrium volume $V$ is at the most 2\%
smaller than the experimental value which correspond to a less than 0.5\%
deviation from the experimental lattice parameter. However the bulk
modulus which is very sensitive to the slop of the total energy versus the
unit cell volume deviates at the most by 10\% in the case of Ge, and  when
calculated at the experimental unit cell volume.  But only by 5\% when
calculated at the theoretical equilibrium volume. Our calculation of the 
bulk modulus is in excellent agreement with other calculation
\cite{meth}.

Figure ~\ref{bandgap} shows  the LDA underestimated (a)  direct band  and (b) minimal gaps 
of Si, Ge, and GaAs compared with the Ge, and GaAs experimental results of
Go\~ni et al. \cite{goni}. For GaAs a cross over from direct band gap to
indirect band gap takes palace between $\Gamma$ and $X$ at approximately 
8 GPa. For Ge this cross over occurs along $\Gamma L$ at a lower pressure
of 3 GPa. The direct band gap increases linearly with pressure and is in 
good agreement with the experimental results for both  GaAs and Ge. There
is no experimental data for Si under hydrostatic pressure.  Table IV present
the first and second order coefficients describing the dependence of the direct band
gap at $\Gamma$ under hydrostatic pressure, 
$E_0 (P) = E_0 + a P + b P^2$, compared to the experimental results of 
Go\~ni et al.\cite{goni}. Apart for the underestimation of the band gap,
the first  and second coefficients of the pressure dependence of the band
gap are in good agreement with the experimental results. This suggest that
the scissors-operator shift is a good approximation for the description of
the band gap under hydrostatic pressure.

\section{Optical properties of  Si, Ge, and GaAs}
\subsection{frequency dependent of the complex dielectric function of Si,
Ge, and GaAs}
Figure ~\ref{eps2si} and ~\ref{eps2gegaas} present the imaginary part of 
the macroscopic dielectric function
of Si, Ge, and GaAs obtained at the experimental ground state  lattice 
parameters except for  Ge where we have compressed the lattice parameter by 
about 1\%. The compression is done  because within LDA 
and at the experimental lattice parameter Ge is a semi-metal.
The LDA $\epsilon_2$ is shifted towards higher energy by the 
scissors-operator  shift
in order that the optical band gap agrees with experiment. The comparison
to experimental results of Aspnes and Studna \cite{aspnes}  
shows that all the features in the experimental spectra are  reproduced by the 
calculation. It is interesting to notice that the calculated LDA 
$\epsilon_2(\omega)$ of Si exhibit the largest
underestimation of the $E_1$ peak  (about 50\% in intensity) whereas, in Ge
and GaAs the underestimation of the $E_1$ peak is only about 12\%. The 
$E_2$ peak is overestimated by LDA by about 34\% for Si, 50\% for Ge, and
60\% for GaAs. This overestimation of the $E_2$ peak by LDA  is due to a 
strong van Hove singularity near the X points of the Brillouin zone where 
parallel bands occur over a large plateau \cite{alouanib,cc}. 
This overestimation can be
reduced substantially by including the life-time broadening of the
quasiparticles through a self-energy calculation.   

The effect of interband electronic transitions due to the  $3d$ semi-core 
states, without scissors operator shift,  is presented in 
Fig. ~\ref{eps2high}a and ~ref{eps2high}b for Ge,
 and GaAs respectively. For 
Ge the onset of transitions begun at photon energy of about 25 eV, and the
intensity is very similar to the $p$-density of states of the empty
states of Ge. This is because the $3d$ states of Ge are very narrow, and
the dipole selection rules allow transitions only to the empty $p$ states of 
Ge, the f-states in this energy range are absent. 
Whereas for GaAs, the onset of transitions begun at 12 eV, and the 
intensity $\epsilon_2$ spectrum above 12 eV is very different from the 
empty $p$ states of Ga. This is because of the relatively large dispersion
of the $3d$ semi-core states of Ga. It should be of interest to confirm
experimentally these theoretical predictions.

The real part $\epsilon_1(\omega)$  of the dielectric function of Si, Ge,
and GaAs calculated by Kramers-Kronig transform of the imaginary part 
$\epsilon_2(\omega)$ are presented in Fig. ~\ref{eps1gegaas} together with the
experimental results of Aspnes and Studna \cite{aspnes}. In the same 
figure we have also presented the scissors-operator shift 
$\epsilon_1^{QP}(\omega)$  and the high
frequency  asymptotic limit $\epsilon_1(\omega) = 1 -  \omega_P^2/
\omega^2$, where $\omega_P$ is the free-electron plasmon frequency. We
notice that the analytic asymptotic limit matches nicely the calculated LDA
$\epsilon_1$, which is an indication of the quality of the calculation. 
For the $\epsilon_1^{QP}$ we need to use a different plasmon frequency as
described in equation (~\ref{intsumruleqp}) due to the poor 
description of the higher excited states by the scissors approximation.
  
In conclusion, we believe that the excitonic effects may be  important 
for the dielectric function of  Si but  less for those of Ge, and  GaAs. 
A QP  calculation of the dielectric function  
including the dynamical screening of the Coulomb interaction, 
like in the GW approximation of Hedin \cite{hedin}, 
would certainly improve the intensity of at
least the   $E_2$ peak by introducing a life-time broadening of the
quasi-particles. 

\subsection{Hydrostatic pressure  dependent of the static  dielectric function 
of Si, Ge, and GaAs}
Figure ~\ref{staticlda} and ~\ref{staticsos} presents the hydrostatic 
pressure dependence of the static
dielectric function, $\epsilon_\infty$,  of Si, Ge, and GaAs calculated
within the LDA without and with the scissors-operator shift (SOS),   
respectively. Our data are
compared to the experimental results of Go\~ni {\it et al.} \cite{goni}
and to the pseudopotential calculations of Si and Ge of Levine and 
Allan \cite{zachary}. Our calculation and the pseudopotential theory of Ref.
\cite{zachary} suggest that LDA is overestimating the static dielectric 
function of Si, Ge, and GaAs over the whole range of hydrostatic pressure, 
and that 
the use of a unique value of the  scissors-operator shift for the correction of the band gap at 
$\Gamma$ produces a nice agreement with the experimental results
\cite{goni}. The static dielectric function decreases almost linearly with
the pressure due to the increase of the direct band gap. However, for
  Si $\epsilon_\infty$ is almost constant with the pressure and this
is because  the increase of the direct band gap  is almost compensated by a 
decrease of the indirect band gap (see Figure \ref{bandgap}).

Table V, VI, and VII present the calculated pressure band
gaps, static dielectric function and f-sum rule for Si, Ge, and GaAs, with
a comparison to the experimental results of Ref. \cite{goni}. The
agreement with the experimental results is excellent when the scissors-operator shift is used.
The f-sum rule deviates at most by 5.2\% from unity in the case of Ge
which reflect the high precision of the calculation of the optical matrix 
elements. The fact that the f-sum rule is not quite exhausted for Ge and 
GaAs (deviation of about 5\%) as compared to Si (deviation of about 1\%)
is not due to a possible incompleteness of our basis set\cite{basisset} but 
rather to our use of
   all electron  electronic structure. When the valence states are very
well isolated from the core states, like in the case of Si where the core
states lie about 80 eV below the valence bands, the sum rule should
be exhausted. However, for Ge and GaAs where the semi-core $3d$-states are
very close to the valence states and greatly affect the optical
properties, the f-sum rule could deviate markedly from unity, i.e.,  the
average effective number of electrons per atom contributing to the optical 
transitions is much larger than 4 electrons per atom \cite{ph63}. In 
pseudopotential theory, since the core states are absent, the f-sum rule
is exhausted for all semiconductors \cite{zachary}. The details of the 
contribution of the $3d$ semi-core states to the oscillator strength and
the study of the effective number of electrons contributing to the optical
transitions  are out of the scope of this paper and will be addressed 
elsewhere. 

The first and second-order coefficients describing the pressure dependence
of the static dielectric function $\epsilon_\infty$ are presented in Table
VIII. The results are compared to the experimental results of Go\~ni et
al. Ref. \cite{goni}, and the pseudopotential calculation of Ref.
\cite{zachary}. The overall agreement with experiment and the
pseudopotential calculation is excellent. 

In Table IX we present our calculation for the static dielectric function
of Si, Ge, and GaAs including the spin-orbit coupling effect at the
variational level, and the effect of the 3$d$ states in the interband 
transitions. The calculated potential 
includes  always the $3d$ states, and only the dielectric function is
calculated with or without the $3d$ interband transitions. We have
obtained that the inclusion of the $3d$ interband transitions increases
slightly the static dielectric function, whereas the spin-orbit coupling 
increases it by 2.1\% and 3.2\% for Ge and GaAs,
respectively. The $\epsilon_\infty$ of
Si is insensitive to the spin-orbit coupling. The calculated scissors-operator 
shift 
$\epsilon_\infty$ including the spin-orbit coupling effect decreases  by about 
3.3\%~ and 4.1\%~ for Ge and GaAs, respectively. This is because the band gaps 
of Ge and GaAs are further reduced in presence of spin-orbit coupling which
resulted in a larger scissors-operator shift for the determination of $\epsilon^{QP}_\infty$.    

\section{Conclusion}
The macroscopic dielectric function in the random-phase-approximation  
without local field effect has been implemented using the local density
 approximation with an all electron, 
full-potential linear muffin-tin
orbital  orbital basis-set. The method is used to calculate  the optical 
properties of the semiconductors, Si, Ge, and GaAs, under hydrostatic
pressure. We have found that the LDA overestimation  
the static dielectric function over all the pressure range from 0 to 12
GPa, and that a single value of the  so called scissors-operator shift 
which  account for the correct band gap at $\Gamma$ produced a good agreement 
with the experimental data of Go\~ni and coworkers \cite{goni}. This makes us
conclude that because LDA underestimates the band gap it is incapable 
of producing the correct static dielectric
function even though $\epsilon_\infty$ is a ground state property.

Since (i) the Kohn-Sham (KS) density functional (DF)
without the local density approximation should in principle  produce the 
correct $\epsilon_\infty$, and (ii) since the LDA calculation with 
the scissors-operator shift also produces the correct $\epsilon_\infty$, we
are tempted to conclude that the  KS-DF theory should produce the
correct band gap for semiconductors.  This conclusion is not confirmed 
by a non-selfconsistent GW calculations which suggest  that the 
true KS-DF theory also underestimates the band gap \cite{gss}.

Our analysis of the dielectric function, the sum rules and the 
energy-loss function shows that while the scissors-operator shift is a good
approximation for the low lying excited states it appears as bad
approximation for the high energy excited states. This is because the 
high energy states are free electron like hence well described within
LDA.

Our calculation of the dynamical dielectric function shows that 
the $E_1$ peak intensity is underestimated for Si by about 50\%, and
for  Ge and GaAs by only 12\%. These results imply that 
the excitonic effects may  be important for the dielectric function of Si,  
but less for those  for  Ge, and  GaAs. 

We have also shown that including the 
$3d$ semi-core states in the interband transitions hardly changes the static 
dielectric function, $\epsilon_\infty$, 
however their contribution to the
intensity of dynamical dielectric function for higher  photon energies 
is substantial, and could be checked experimentally.  We have also found
that  the spin-orbit coupling 
has a significant effect on $\epsilon_\infty$ of Ge and GaAs, but not of
Si.

We  thank J. W. Wilkins for interesting discussions. 
This research was supported in part by the U.S. Department of Energy 
Basic Energy Sciences,
Division of Materials Sciences and by NSF, grant number DMR-9520319.
Supercomputer time was provided by the Ohio State Supercomputer Center.

\newpage
\begin{figure}
\caption{ 
\label{energyloss}
Calculated energy-loss function of GaAs within the LDA (full curve) and
within the scissors approximation (dashed curve). It is clearly seen that
the maximum of the LDA curve is much closer to the free valence electron 
plasma frequency of 15.5 eV. 
}
\end{figure}
\begin{figure}
\caption{ 
\label{bandgap}
Calculated (a) direct band gap E$_0$, and (b) minimum band gap $E_{\rm gap}$ 
 of Si, Ge, and GaAs as a function of 
hydrostatic pressure compared to the experimental results of Go\~ni 
{\it et al.} \protect\cite{goni} for Ge (dashed line), and GaAs (thick
line). 
Plot (a) shows that the direct band gaps increase almost linearly with 
pressure. Plot (b) shows that for  GaAs there is a cross over of the band gap
from direct to indirect at around 8 GPa, and a cross over for Ge at almost
3 GPa. The indirect band gap of Si decreases linearly with pressure. 
}
\end{figure}
\begin{figure}
\caption{Calculated imaginary part of the dielectric function of Si 
at the experimental equilibrium volume, shifted by $\Delta = 0.6$ eV
towards  higher photon energies,  compared with the experimental
results of  Ref. \protect\cite{aspnes}. The experimental $E_1$ structure at
4 eV is underestimated whereas the  main $E_2$ structure at 4.5 eV is
overestimated.  
\label{eps2si}}
\end{figure}
\begin{figure}
\caption{Imaginary part  of the dielectric function of Ge at 10 kbar shift
by 0.4 eV and GaAs at the experimental equilibrium volume shifted 
by 1.1 eV, compared with the experimental results of 
Ref. \protect\cite{aspnes}. In both Ge and GaAs $\epsilon_2 (\omega)$ the 
experimental $E_1$ is only slightly underestimated and $E_2$ is
overestimated. 
\label{eps2gegaas}}
\end{figure}
\begin{figure}
\caption{Contribution of the $3d$ interband transitions to the imaginary part  
of the dynamical dielectric function of (a) Ge (at 10 kbar) and (b) GaAs
at the experimental equilibrium volume. The full line and the dashed line 
are with and without $3d$ interband transitions, respectively. Due to the
narrow nature of the $3d$ semi-core states of  Ge, 
the intensity of $\epsilon_2$ above 25 eV is very similar to the
empty $p$-density of states of Ge. Whereas for GaAs, the $3d$ semi-core
states of Ga are relatively delocalized, which makes the intensity of
$\epsilon_2$ above 12 eV much different from the Ga empty $p$-density
of states. 
\label{eps2high}}
\end{figure}
\begin{figure}
\caption{Real  part  of the dielectric function of Ge (at 10 kbar) and GaAs,
at the experimental equilibrium volume, compared with the experimental
results of Ref. \protect\cite{aspnes}. The analytic asymptotic limit, 
shown by the empty circles, matches nicely the calculated spectra
above 10 eV. 
\label{eps1gegaas}}
\end{figure}
\begin{figure}
\caption{LDA scalar-relativistic  calculated static dielectric function of Si, 
Ge, and GaAs as a function of
hydrostatic pressure compared to the experimental results of Go\~ni 
{\it et al.} \protect\cite{goni} and the pseudopotential calculation of
Levine and Allan\protect\cite{zachary}.  
\label{staticlda}}
\end{figure}
\begin{figure}
\caption{ LDA plus the scissors-operator shift (SOS) calculated static 
dielectric function of Si, Ge, and GaAs as a function of
hydrostatic pressure compared to the experimental results of Go\~ni 
{\it et al.} \protect\cite{goni}.  
\label{staticsos}}
\end{figure}
\widetext
\begin{table}
\caption{
Basis-sets used for the calculation of the excited states of 
Si, Ge and GaAs. Each orbital has different kinetic energy $\kappa^2$ 
in its the interstitial region. For example, the $3s$
orbital of Si is used three times, each of the $3s$ wave function has a
different kinetic energy in the interstitial region.}
\begin{tabular}{cc}
 { Si}: & $3 \times (3s,3p)$     \hskip 0.5truecm $2 \times (3d) $       \\
           & $3 \times (4s,4p)$     \hskip 0.5truecm $2 \times (4d,4f)$     \\
 { Ge}: &                         \hskip 0.5truecm $2 \times (3d)$       \\
           & $3 \times (4s,4p)$     \hskip 0.5truecm  $2 \times (4d) $       \\
           & $2 \times (5s,5p)$                             \\
{ GaAs}:&                          $2 \times (Ga\ 3d)$    \\
           & $3 \times (Ga\ 4s,4p)$ \hskip 0.5truecm  $2 \times (Ga\ 4d) $   \\
           & $3 \times (Ga\ 5s,5p)$                         \\
           & $3 \times (As\ 4s,4p)$ \hskip 0.5truecm  $2 \times (As\ 4d) $   \\
           & $3 \times (As\ 5s,5p)$                         \\
\end{tabular}
\label{table1}
\end{table}
\begin{table}
\caption{
Eigenvalues of Si at high symmetries points ($\Gamma$, $X$, and $L$) as 
compared to the results produced by means of a  linear combination of  
gaussian orbitals \protect\cite{germans},
and by pseudopotential (PP) method \protect\cite{cvl}. The zero
of energy is chosen at $\Gamma_{25v^\prime}$ point.
}
\begin{tabular}{cccc}
High-symmetry & Gaussian & PP    & Present     \\
point         & orbitals &       & Calculation \\
\noalign{\hrule}
$\Gamma_{1v}         $&$ -11.91 $&$    -11.91 $&$   -11.96 $\\
$\Gamma_{25v^\prime} $&$ 0.0    $&$    0.0    $&$   0       $\\
$\Gamma_{15c}        $&$ 2.57   $&$    2.55   $&$   2.56   $\\
$\Gamma_{2c^\prime}  $&$ 3.24   $&$    3.28   $&$   3.20   $\\ 
                      &           &             &             \\
$X_{1v}             $ & $ -7.77  $&$    -7.76  $&$   -7.82  $\\
$X_{4v}          $    & $ -2.78  $&$    -2.86  $&$   -2.83  $\\
$X_{1c}          $    & $ 0.65   $&$    0.66   $&$   0.62   $\\
$X_{4c}          $    & $ 10.03  $&          &$   10.03  $\\
                      &           &             &             \\
$L_{2v^\prime}   $    & $ -9.58  $&$    -9.56  $&$   -9.63  $\\
$L_{1v}          $    & $ -6.94  $&$    -6.96  $&$   -6.99  $\\
$L_{3v^\prime}   $    & $ -1.17  $&$    -1.20  $&$   -1.19  $\\
$L_{1c}          $    & $ 1.47   $&$    1.50   $&$   1.44   $\\
$L_{3c}          $    & $ 3.32   $&$    3.33   $&$   3.31   $\\
$L_{2c^\prime}   $    & $ 7.77   $&$           $&$   7.66   $\\
\noalign{\hrule}
 Indirect band gap          &$ 0.52   $ &$           $&$   0.50   $\\
\end{tabular}
\label{table2}
\end{table}
\begin{table}
\caption{
Calculated equilibrium volume (V), electronic pressure, and bulk modulus
of Si, Ge, and GaAs.
The the bulk modulus are calculated both at the experimental 
($V_0$) and  theoretical (V) unit cell volumes. The experimental results 
are shown between parenthesis.
}
\begin{tabular}{cccccc}
Semiconductor    & V$_0$     & V/V$_0$ & P(V$_0$) & B(V)   & B(V$_0$) \\
               & (\AA$^3$)  &          & (GPa)     & (GPa)   & (GPa)    \\
\noalign{\hrule}
   { Si}    & $39.98$   & $.990$   & $-.70$    & $95.8$ &  $91.2$   \\
               &            &          &           &   &$(98.8)$  \\
   { Ge}    & $45.27$   & $.988$   & $-.80$    & $71.0$  & $67.1$   \\
               &            &          &           &  &$(74.4)$  \\
   { GaAs}  & $45.12$   & $.984$   & $-1.2$    & $74.2$  & $69.3$   \\
               &            &          &           &  &$(74.7)$  \\
\end{tabular}
\label{table3}
\end{table}
\begin{table}
\caption{
First and second-order coefficients describing the dependence of the
direct band gap  at $\Gamma$ ($E_0$) under  hydrostatic pressure 
($E_0 (P) = E_0 + a P + b P^2$) for Si, Ge, and GaAs. 
The experimental results are
from Go\~ni et al. Ref. \protect\cite{goni}.
 }
\begin{tabular}{lcccccr}
 &  \multicolumn{2}{c}{$E_0$} & \multicolumn{2}{c} {a (meV/GPa) }
& \multicolumn{2}{c} {b (meV/GPa$^2$)} \\ 
        & Theory  &Expt.    & Theory & Expt.  & Theory & Expt. \\
Si      &  3.273  &         & 100.8  &        & 0.05   &       \\
Ge      & -0.084  & 0.795   & 125.4  & 121    & 0.2    & 0.2   \\
GaAs    &  0.41   & 1.43    &  99.1  & 108    &-0.1    & -0.1  \\
\end{tabular}
\label{table4}
\end{table}
\begin{table}
\caption{
Calculated pressure, band gap, static dielectric function with and
without scissors-operator shift (SOS), and the f-sum rule
of Si as a function of volume. 
 \protect\cite{germans},
}
\begin{tabular}{ccccccc}
 V$_0$/V & P(GPa) & f-sum & gap(eV) & $\epsilon_\infty$ (LDA)& 
$\epsilon_\infty$ (SOS) & $\epsilon_\infty$ (Expt.) \\
1.000 & -.75 & 0.988 & 0.50 & 13.75 & 12.08& 12.0$^{\rm a}$\\
1.025 &  1.8 & 0.989 & 0.46 & 13.65 & 12.00& \\
1.050 &  4.8 & 0.987 & 0.42 & 13.61 & 11.96& \\
1.100 &  9.8 & 0.991 & 0.34 & 12.57 & 11.98& \\
\end{tabular}
$^{\rm a}$ Ref.~\protect\cite{harrison}
\label{table5}
\end{table}
\begin{table}
\caption{
Calculated pressure, band gap, static dielectric function with and
without scissors-operator shift (SOS), and the f-sum rule
of Ge as a function of volume. The experimental data are from Go\~ni et
al. \protect\cite{goni}.  }
\begin{tabular}{ccccccc}
 V$_0$/V & P(GPa) & f-sum & gap(eV) & $\epsilon_\infty$ (LDA)& 
$\epsilon_\infty$ (SOS) & $\epsilon_\infty$ (Expt.) \\
1.025    & 1.0    &  1.053 &  .04    & 19.24 & 15.32 & 15.59\\
1.050    & 2.9    &  1.041 &  .21    & 18.14 & 14.71 & 15.01\\
1.075    & 5.0    &  1.042 &  .29    & 17.44 & 14.33 & 14.49\\
1.100    & 7.2    &  1.043 &  .37    & 16.70 & 13.90 & 14.07\\
1.150    & 12.3   &  1.045 &  .48    & 15.81 & 13.46 & 13.63\\
\end{tabular}
\label{table6}
\end{table}
\begin{table}
\caption{
Calculated pressure, band gap, static dielectric function with and
without scissors-operator shift (SOS), and the f-sum rule
of GaAs as a function of volume. The experimental data are from Go\~ni et
al. \protect\cite{goni}.  }
\begin{tabular}{ccccccc}
 V$_0$/V & P (GPa) & f-sum & gap(eV) & $\epsilon_\infty$ (LDA)& 
$\epsilon_\infty$ (SOS) & $\epsilon_\infty$ (Expt.) \\
1.000    & -1.2   & 1.041  & .29     & 14.44 & 11.0  & 11.05\\
1.025    &  .68   & 1.042  & .48     & 13.93 & 10.72 & 10.88\\
1.050    &  2.8   & 1.043  & .66     & 13.45 & 10.53 & 10.69\\
1.075    &  4.7   & 1.044  & .85     & 13.09 & 10.41 & 10.53\\
1.100    &  6.8   & 1.044  & 1.02    & 12.75 & 10.25 & 10.34\\
1.150    &  11.8  & 1.046  & 1.08    & 12.20 & 10.03 &  9.90\\
\end{tabular}
\label{table7}
\end{table}
\begin{table}
\caption{
First and second-order coefficients describing the dependence of the
static dielectric function on hydrostatic pressure ($\epsilon_\infty (P)
= \epsilon^0_\infty + a P + b P^2 $) for Si, Ge, and GaAs. The experimental 
data are from Go\~ni et
al. \protect\cite{goni}.  }
\begin{tabular}{lcccccccr}
 &\multicolumn{2}{c}{$\epsilon^0_\infty$} &\multicolumn{2}{c}{a(1/GPa)} 
& \multicolumn{2}{c}{b (1/GPa$^2$)} & \multicolumn{2}{c}
{$d\ln (\epsilon_\infty) / d P (10^{-12}$/{\rm Pa})} \\ 
      & Theory &Expt.   & Theory &Expt.           & Theory & Expt. & Theory &
 Expt.\\ 
Si    &  12.05 &        & -0.032 &                & 0.0025 &                &
 -2.65  &      \\
 Si   &  11.16$^{\rm a}$&        &-0.027$^{\rm a}$&        &0.0013$^{\rm a}$& 
      &-2.6,-2.43$^{\rm a}$   &   \\
Ge    &  15.58          & 15.94  & -0.32          & -0.36  & 0.012          & 
0.014 & -20.21& -22.60 \\
      &  16.04$^{\rm a}$ &      & -0.46$^{\rm a}$ &        & 0.018$^{\rm a}$ &    & -31,28.66$^{\rm a}$  &   \\
GaAs  &  10.83 &10.92     & -0.11 & -0.09  & 0.004 &       & -10.43 &-8.06 \\
\end{tabular}
\label{table8}
$^{\rm a}$ Pseudopotential calculation of Ref.\protect\cite{zachary},  slightly
larger numbers are quoted for $d\ln(\epsilon_\infty)/dP$ in their Table VI.  
\end{table}
\begin{table}
\caption{
Calculated static dielectric function of Si, Ge, and GaAs at the
equilibrium lattice parameter (except for Ge where it is calculated at 
a slightly smaller lattice parameter (1\% smaller)  than the experimental one
because Ge is a metal in LDA for $V/V_0$ = 1.). The calculation are
done using scalar relativistic (SR) LMTO without 3$d$ states, with the 
$3d$ states (SR+3$d$), with the spin-orbit coupling at the variational
level (SR+SO), and with the SO coupling and the 3$d$ states included
(SR+SO+3$d$). 
  }
\begin{tabular}{lcccccc}
 &  \multicolumn{2}{c}{Si} & \multicolumn{2}{c}{Ge}
& \multicolumn{2}{c}{GaAs} \\ 
 & LDA&LDA+SOS.& LDA & LDA+SOS & LDA & LDA+SOS \\
SR       & 13.75 &  12.08& 18.14 &  14.71  & 14.44&    11.0  \\
SR+3$d$  &       &       & 18.16 &  14.73  & 14.47&  11.03   \\
SR+SO    &  13.69& 12.0  & 18.52 & 14.23   & 14.90& 10.52    \\
SR+SO+3$d$&       &       & 18.54 & 14.25   & 14.93& 10.55    \\
Expt.    &       &12.0$^{\rm a}$ 11.4$^{\rm b}$   &       & 14.98$^{\rm c}$   &      & 
10.9$^{\rm a}$   \\
\end{tabular}
\label{table9}
$^{\rm a}$ Ref.~\protect\cite{harrison}
~~~$^{\rm b}$ Ref.~\protect\cite{li}
~~~$^{\rm c}$ Ref.~\protect\cite{goni}
\end{table}

\begin{references}
\bibitem{commenta}Here the principal quantum number $\nu$ and the spin channel 
$\sigma$ are ignored for simplicity, however
our numerical code has the spin polarization capability, and the use of 
the different $\nu$ with the same momentum channel in the basis set. 
We use also only
one $\kappa$ variational parameter per orbital (see definition in the
text) in all formulas to avoid 
using it as an obvious index to the wave function and  extra 
summation in the Bloch wave function.
\bibitem{basisset} We have checked carefully the basis set by increasing
the number of
Hankel functions for negative kinetic energies and Neumann functions for
positive kinetic energies $\kappa^2$ and have not been able to improve
on the f-sum rule. Our basis-set indicates that all physical properties
quantities  have been converged, including the total energy which is
converged at the tenth of the mRy level.
\end{references}
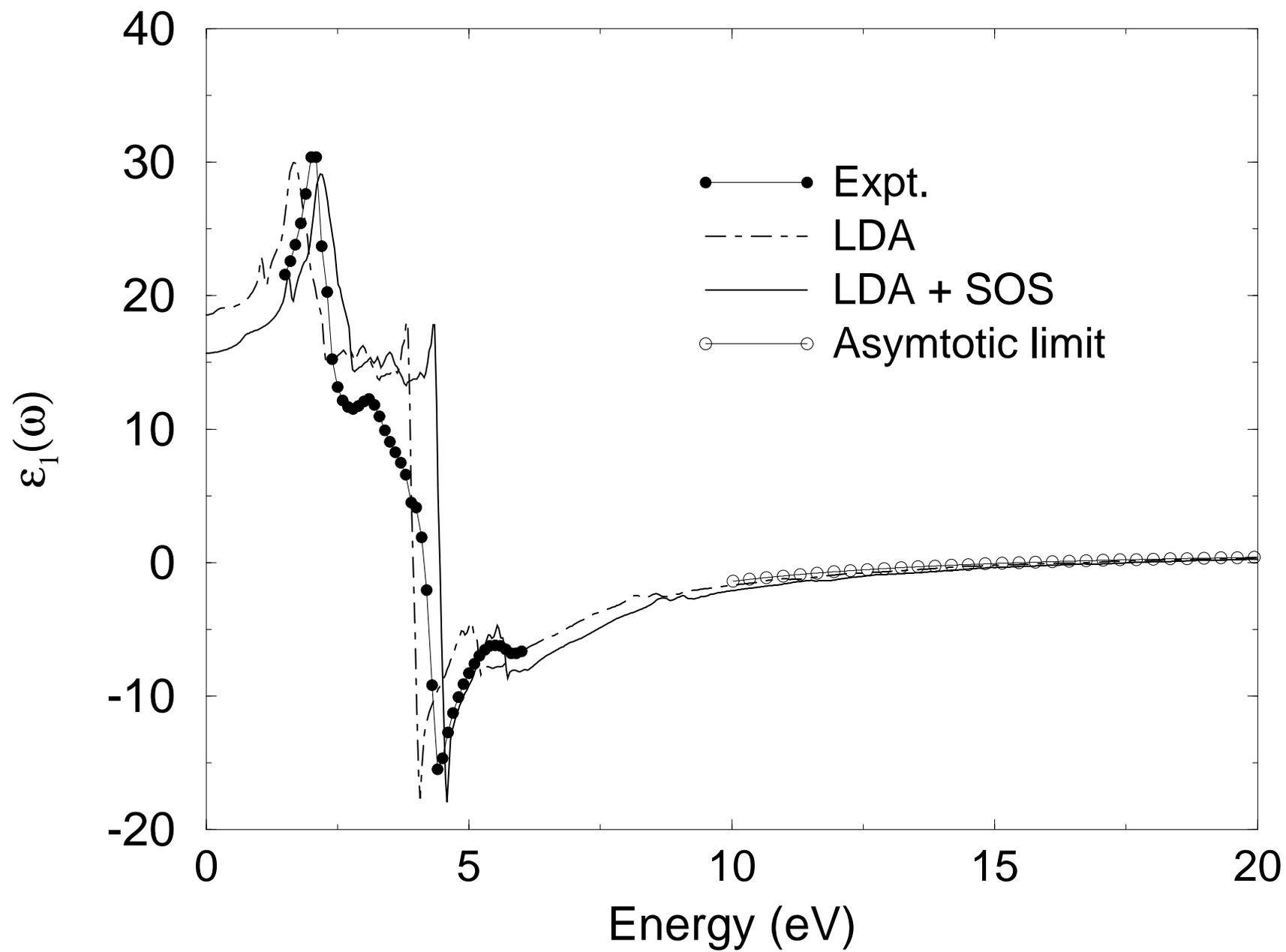
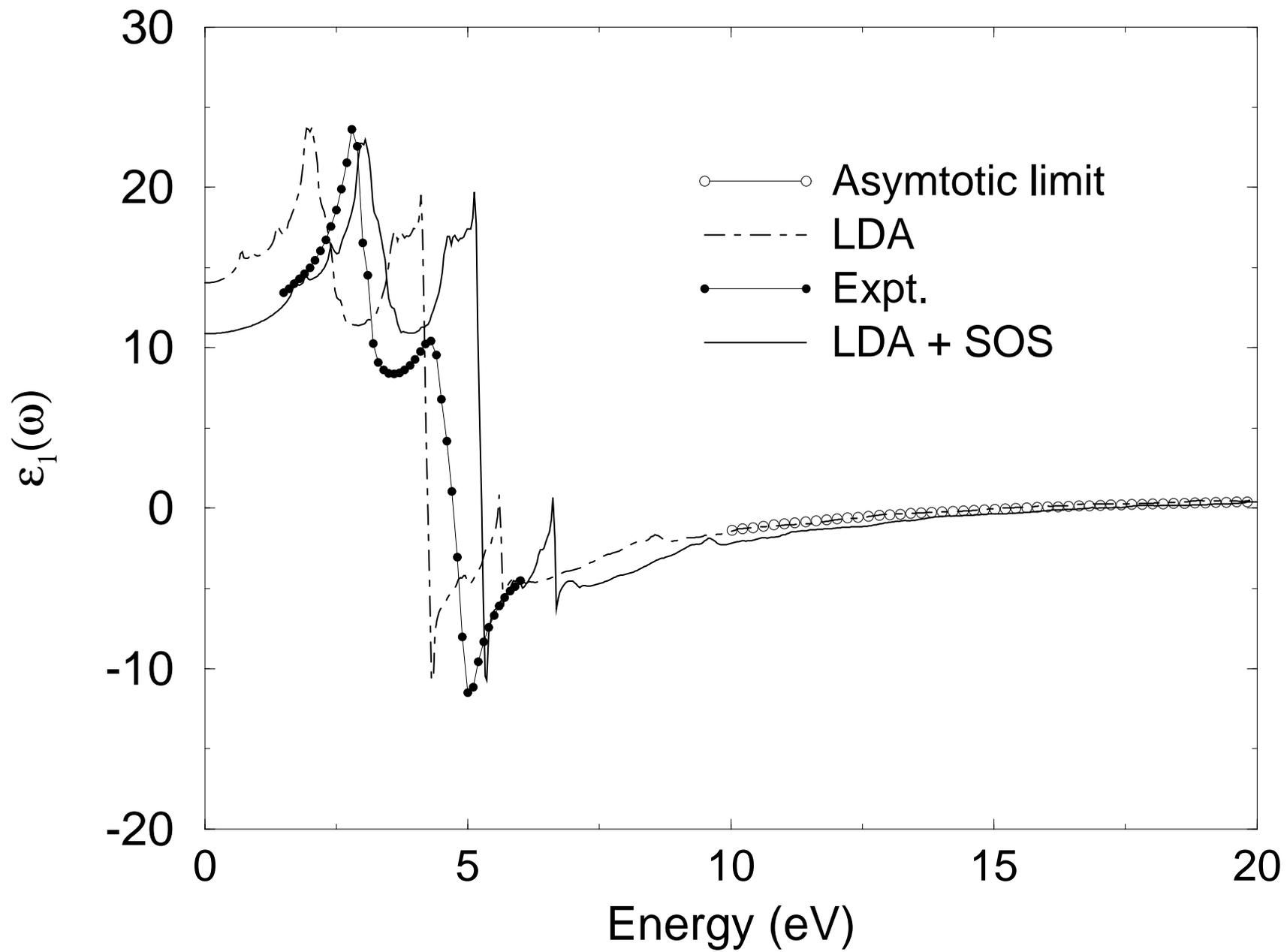
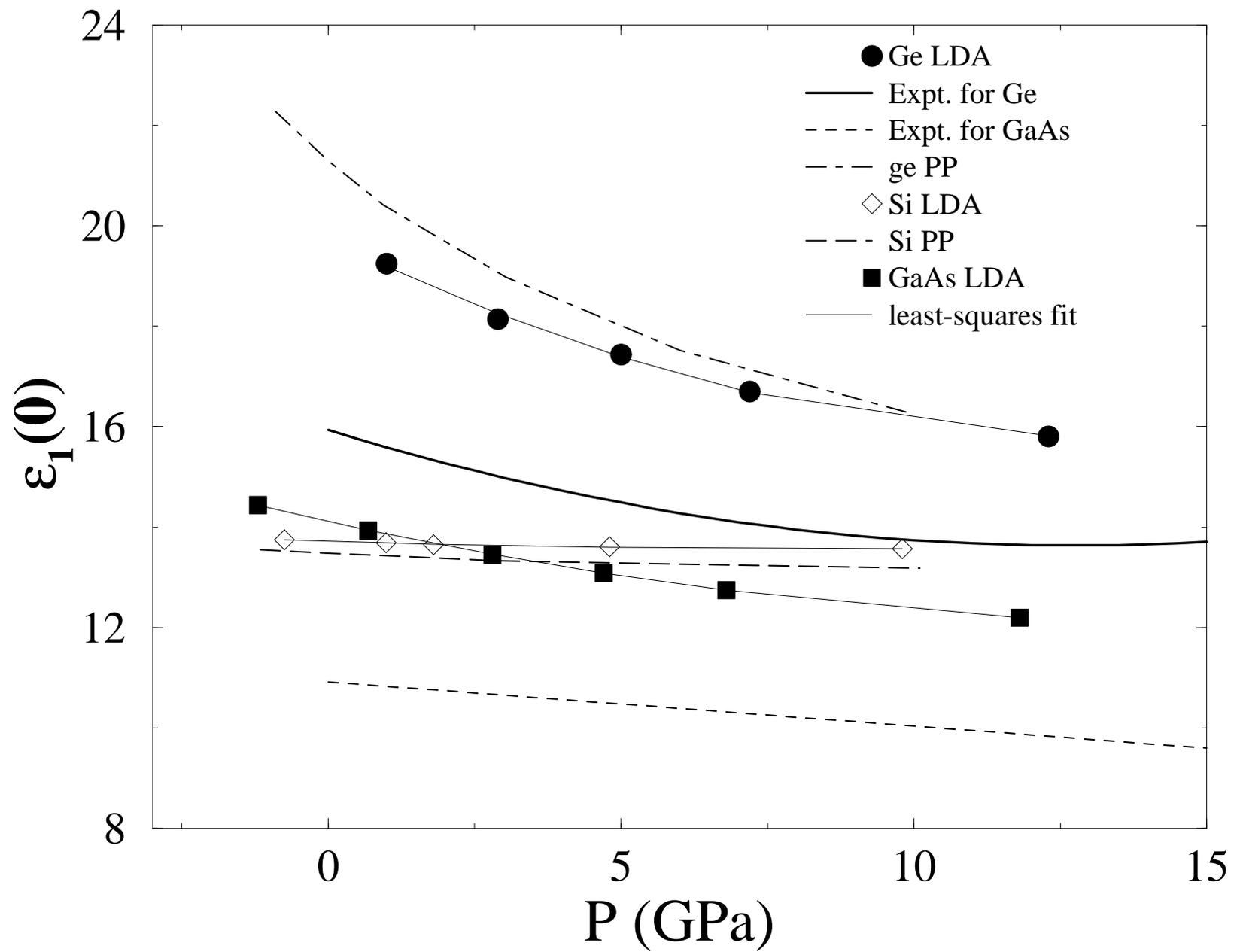
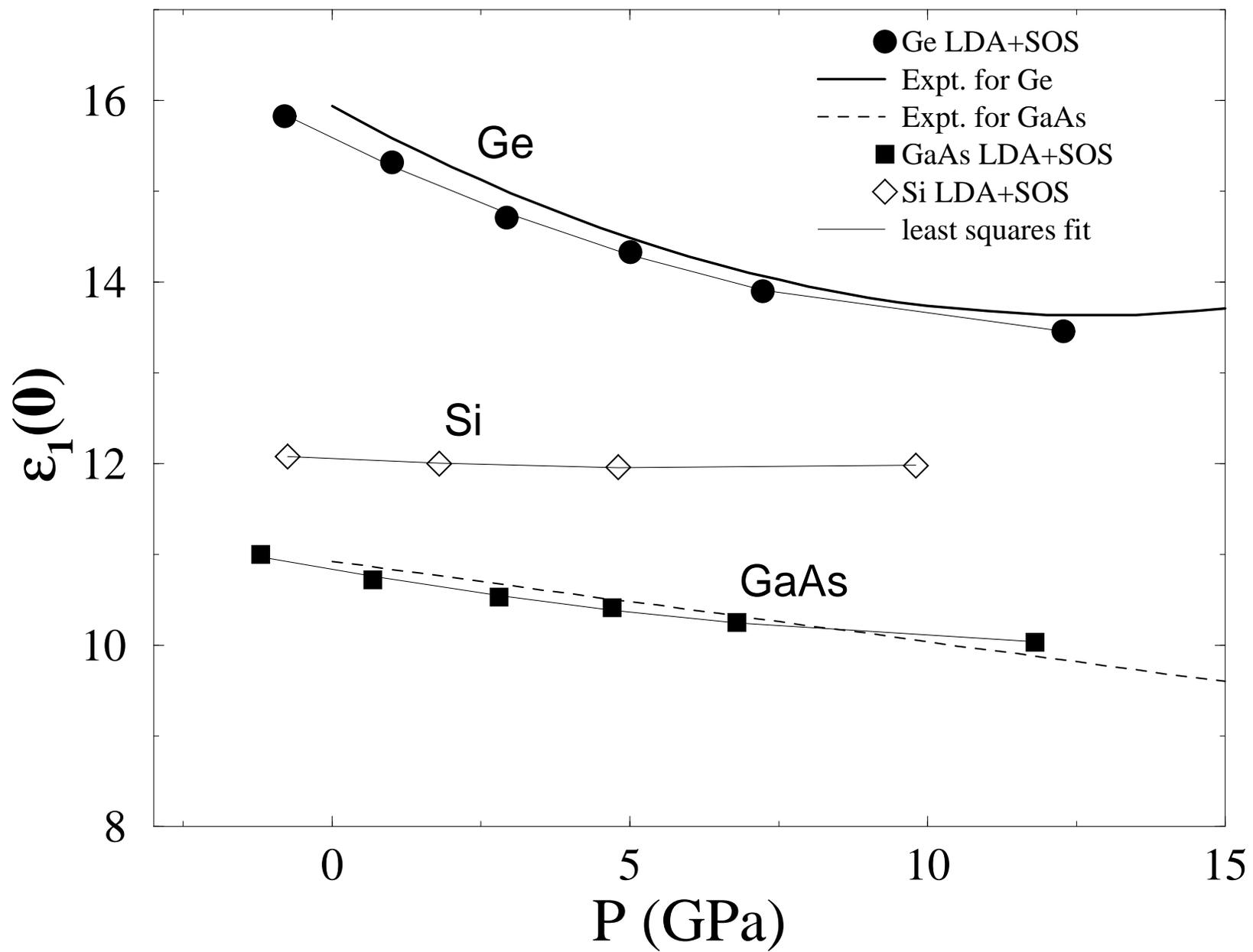
\end{document}